\documentclass[aps,floatfix,preprint,nofootinbib]{revtex4}

\usepackage{graphicx}
\usepackage{epic}
\usepackage{eepic}
\usepackage{latexsym}

\newcommand{\eq}[1]{(\ref{#1})}
\newcommand{\be}{\begin{equation}}
\newcommand{\ee}{\end{equation}}
\newcommand{\bea}{\begin{eqnarray}}
\newcommand{\eea}{\end{eqnarray}}

\newcommand{\vs}[1]{\vspace{#1 mm}}

\newcommand{\bk}{{\bf k}}
\newcommand{\bx}{{\bf x}}
\newcommand{\bkp}{{\bf k'}}

\newcommand{\bJ}{{\bf J}}
\newcommand{\bK}{{\bf K}}
\newcommand{\bn}{{\bf n}}
\newcommand{\bmm}{{\bf m}}

\def\a{\alpha}
\def\b{\beta}

\def\d{\delta}
\def\D{\Delta}
\def\e{\epsilon}

\def\fr{\frac}

\def\o{\omega}
\def\dg{\dagger}

\let\bm=\bibitem
\def\nn{\nonumber}

\begin{document}

\title{Cosmological particle creation:\\  Fluctuations and an ensemble picture}

\author{Ali Kaya}
\email[]{ali.kaya@boun.edu.tr}
\affiliation{Bo\~{g}azi\c{c}i University, Department of Physics, \\ 34342,
Bebek, \.Istanbul, Turkey \vs{3} \\ and  \vs{3}\\ Feza G\"{u}rsey Institute,\\
Emek Mah. No:68, \c{C}engelk\"{o}y, \.Istanbul, Turkey\vs{3}}

\date{\today}

\begin{abstract}

We point out that in the context of quantum fields in time dependent classical backgrounds, the number of created particles with a given momentum largely deviates about its mean value. Since the corresponding Fourier modes are nonlocal, this deviation shows that the expectation value of the number operator can only make sense in an ensemble of spacetimes. Using a complete orthonormal family of localized wave packets, we show how an ensemble interpretation can be given to cosmological particle creation in local terms. The reheating process following inflation is reexamined in the light of this construction. 

\end{abstract}

\maketitle

One of the main assumptions of quantum theory is that a physical quantity is well defined in a state if the state is an eigenvector of the operator corresponding to that quantity. When this is not the case, the quantity does not acquire a definite value and one can only talk about the probability of seeing an outcome of a measurement which can be understood in an ensemble picture. The expectation value of the corresponding operator gives the average of the  measurements done on the identical copies of the same system in the ensemble. The aim of this work is to reinterpret cosmological particle creation in time dependent classical backgrounds by taking into account this basic ensemble picture and discuss possible implications of this reinterpretation for inflation. 

To illustrate our point, let us start by reviewing the well known quantum mechanical harmonic oscillator with externally fixed time dependent frequency $\o$. This simple system essentially mimics the basic features of cosmological particle creation. Similar to the usual harmonic oscillator, by introducing suitable {\it time varying} ladder operators  $a$ and $a^\dagger$ the Hamiltonian can be brought into the form $H=\omega\,\left[a^\dagger a +1/2\right]$ (see e.g. \cite{r1}). Therefore, the instantaneous ground state of the system at time $t$ can be defined by $a|0_t>=0$ and a complete orthonormal basis of energy eigenvectors can be introduced as $|n>=\left[(a^\dg)^n/\sqrt{n!}\right]|0_t>$. 

Time dependence of the ladder operators can be solved by a Bogoligov transformation 
\bea
a&=&\a\,a_0 + \b^*\, a_0^\dagger,\nn\\
a^\dagger&=& \b\,a_0+ \a^*\, a_0^\dagger, \label{bog}
\eea
where $a_0$ and $a_0^\dagger$ are the constant operators at time $t_0$ and $\a,\b$ are complex functions of time with $|\a|^2-|\b|^2=1$. It immediately follows from this construction that the ground state of the system at time $t_0$, i.e.  $a_0|0_0>=0$, will not remain to be the ground state at a later time. Instead, the expectation value of the number operator $N=a^\dagger a$ can be found as $<N>=<0_0|N|0_0>=|\b|^2$, which shows that {\it on the average} this state contains $|\b|^2$ quanta at time $t$.

Since $|0_0>$ is {\it not} an eigenvector of the number operator $N$, the above average can only be understood in an ensemble picture. Therefore, it is crucial to determine the deviations about the mean value. If there is a sharp maximum around the mean with a small deviation, then to a very good approximation the ensemble interpretation can be ignored. Expanding $|0_0>$ in the basis vectors $|n>$ and using \eq{bog}, one can determine the probability $P_{2n}$ of finding $2n$ number of quanta in $|0_0>$ as
\be\label{pd}
P_{2n}=\fr{(2n)!}{2^{2n}(n!)^2}\fr{|\b|^{2n}}{|\a|^{2n+1}}. 
\ee
Unfortunately, this distribution does not have a maximum around the average $|\b|^2$. On the contrary, it is a decreasing function of $n$. This shows that although on the average $|0_0>$ contains $|\b|^2$ quanta, {\it the most probable} outcome of a single measurement is the ground state $|0_t>$ with no quanta. Not surprisingly, the deviation $\D N$ about the average is large. Using $(\D N)^2\equiv<N^2>-<N>^2$ one can calculate the relative deviation as
\be\label{dn}
\fr{\D N}{<N>}=\sqrt{2}\,\fr{|\a|}{|\b|}>\sqrt{2},
\ee
which shows that the statistical fluctuations about the mean is quite big.

Let us now consider a free, massive, real scalar field on a cosmological Robertson-Walker background $ds^2=-dt^2+a^2(dx^2+dy^2+dz^2)$. One may assume that in addition to the scale factor $a$, the mass parameter $M$ may also depend on $t$ so that particle creation can arise due to a  coupling to an external time varying field as in the case of reheating. Expanding the field in Fourier modes, one can see that this free field theory is nothing but an infinite collection of harmonic oscillators each of which is labeled by a comoving wave vector $\bk$ and a time dependent frequency  $\omega_k^2=M^2+k^2/a^2-9H^2/4 -3\dot{H}/2$, where $H=\dot{a}/a$ is the Hubble parameter (see e.g. \cite{r2}). As in the harmonic oscillator, one can introduce time dependent ladder operators $a_\bk$ and $a_\bk^\dg$ obeying $[a_\bk,a^\dg_\bkp]=\d(\bk-\bkp)$, and construct the usual Fock space on the instantaneous ground state defined by $a_\bk |0_t>=0$.

The particle creation effects in this setup can also be studied by introducing a Bogoligov transformation 
\bea
a_\bk&=&\a_k a_\bk(t_0) +\b_k^* a_{-\bk}^\dg(t_0),\nn\\
a_{-\bk}^\dg&=&\b_k a_\bk(t_0) +\a_k^* a_{-\bk}^\dg(t_0),
\eea
where the time dependent complex coefficients satisfy $|\a_k|^2-|\b_k|^2=1$ and the modes having momenta $\pm\bk$ are coupled because of the reality of the field. As before the ground state of the system at time $t_0$, defined by $a_{\bk}(t_0)|0_0>=0$, becomes a multiparticle state at a later time.

To find the mean number of created particles in the mode $\bk$, one can calculate the expectation value of the number operator $N_\bk=a^\dg_\bk a_\bk$, which reads 
\be\label{fn}
<N_\bk>=<0_0|N_\bk|0_0>=|\b_k|^2 \d({\bf 0}).
\ee
The singularity in the delta function can be explained as $\d({\bf 0})=V/(2\pi)^3$, where $V$ is the total comoving volume of the space. Defining the comoving number density as $n_\bk=N_\bk/V$ one obtains 
\be\label{nn}
<n_\bk>=<0_0|n_\bk|0_0>=\fr{|\b_k|^2}{(2\pi)^3},
\ee
which is the familiar expression used in the literature (see e.g. \cite{r3}). 

How one should interpret \eq{nn} in quantum theory? Since the mode $\bk$ is globally defined and $|0_0>$ is not an eigenvector of $n_\bk$, \eq{nn} only gives the average of observations carried out  in {\it an ensemble of spacetimes}.  Due to the large deviation about the mean (as in the harmonic oscillator discussed above), it is meaningless to use the expectation value \eq{nn} in a single realization. The situation is different for the total number density $n=\int\, d^3k\, n_{\bk}$ or for the number density operator integrated out in a finite momentum domain as one can see that in both cases the relative deviation in the state $|0_0>$ drops like $1/\sqrt{V}$, which vanishes in the infinite volume limit. However, in the strict limit the momentum modes are completely dislocalized and the density concept becomes ill defined. 

It is clear that the above mentioned problem with \eq{nn} arises due to the nonlocal nature of Fourier modes. To reinterpret particle creation in local terms,  we introduce a new set of ladder operators corresponding to localized wave packets \cite{r4} as follows. Let $\e>0$ be an arbitrary comoving momentum scale and introduce two vectors $\bJ=(j_1,j_2,j_3)$ and $\bn=(n_1,n_2,n_3)$ with {\it integer} entries. Let 
\be\label{ajn}
a_{\bJ\bn}= \fr{1}{\e^{3/2}}\int_{\bJ}\,\exp(-2\pi i\bn .\bk/\e)\,a_\bk,
\ee
where we use a shorthand notation for a three dimensional momentum integral  
\be\label{di}
\int_{\bJ}\equiv \int_{j_1\e}^{(j_1+1)\e} \int_{j_2\e}^{(j_2+1)\e}\int_{j_3\e}^{(j_3+1)\e}\,dk_1 dk_2 dk_3. 
\ee
These operators obey $[a_{\bJ\bn},a^\dg_{\bK\bmm}]=\d_{\bJ\bK}\d_{\bn\bmm}$ and it is easy to see that $a_{\bJ\bn}$ is the annihilation operator for the mode function 
\be\label{mf}
f_{\bJ\bn}({\bf x})=\fr{1}{(2\pi\e)^{3/2}}\int_\bJ \exp(-i\bk.({\bf x}-2\pi\bn/\e)),
\ee
which is peaked around the comoving position $2\pi\bn/\e$ with a spread $1/\e$ in each direction. These localized wave packets form a complete orthonormal family and thus a legitimate basis in the Hilbert space, as good as the Fourier modes $(2\pi)^{-3/2}\exp(i\bk .\bx)$. Roughly speaking, \eq{mf} can be viewed as a mode function of a particle localized in a cubic volume of side length $2\pi/\e$, carrying an average comoving momentum $\bJ \e$.

The cosmological production of these excitations can be determined from the expectation value of the number operator $N_{\bJ\bn}=a^\dg_{\bJ\bn}a_{\bJ\bn}$. The number density $n_{\bJ\bn}$ is defined by dividing $N_{\bJ\bn}$  to the volume of the cube $(2\pi/\e)^3$. The expectation value of $n_{\bJ\bn}$ in the state $|0_0>$ can be calculated as 
\be\label{nn2}
< n_{\bJ\bn}>=<0_0| n_{\bJ\bn} |0_0>=\fr{1}{(2\pi)^3}\int_\bJ |\b_k|^2. 
\ee
The $\e$ dependence of this expression is hidden in the definition of the momentum integral \eq{di} and due to the underlying translational invariance it  does not depend on $\bn$. Contrary to \eq{nn}, it is possible to interpret \eq{nn2} even in a single spacetime as it gives the average of independent observations carried out in different spatial cubic regions labeled by $\bn$. Namely, there is no difference in taking the average over the same cube in different spacetimes or over different cubes of the same spacetime. In this picture the locality of the  particle creation process is evident. 

Consider now the the expectation value of the total number density $n_\bn=\sum_\bJ n_{\bJ\bn}$, which determines the average number of excitations in each cubic region.  Not surprisingly, $<n_\bn>$ is equal to $<n>$ defined in the paragraph below \eq{nn}, which shows that both pictures agree on the average number of created particles. However, instead of decreasing like $1/\sqrt{V}$, the relative deviation of $n_\bn$ in the state $|0_0>$ behaves differently as $\left(\D n_\bn/<n_\bn>\right)\sim \e^{3/2}$ \cite{r5}.\footnote{One simple way to see this dependence is to note that in a homogeneous background the number of particles in a volume $1/\e^{3}$ is proportional to $1/\e^{3}$, and the relative deviation should decrease with the square root of the number of degrees of freedom.} Therefore, even in the infinite volume limit the number of excitations varies from one cubic region to another. Although the background is homogeneous, there appears inhomogeneities due to the quantum mechanical fluctuations and locality of the particle creation process. 

For small $\e$, the volume of the cubic regions that are observed becomes large and the deviation in the number density naturally diminishes. For large $\e$ the opposite is true. In an experimental setup $\e$ should be determined by the details of the measurement. On the other hand in an expanding universe the Hubble radius sets a natural scale since by causality particle creation effects should take place independently in each Hubble volume and local interactions cannot restore homogeneity. 

As an illustrative application of the above formalism, let us finally consider the reheating process after inflation in the parametric resonance regime. Assume that particles are produced around a  momentum $k$ in a band having a width $\D k$ with equal average production. For each wave packet in the band, the deviation in the number of created quanta is approximately given by the quantum mechanical result \eq{dn}. Since the momenta are split up by $\e$, there are nearly $k^2 \D k/\e^3$ number of different wave packets in this band. Therefore, the relative deviation of the total number of created particles should behave like $\left(k^2 \D k/\e^3\right)^{-1/2}$ (note the $\e$ dependence). As shown in \cite{r6}, in a simple chaotic model with the inflaton potential $m^2\phi^2$, scalar $\chi$ particles can be produced in the parametric resonance regime due to the coupling $-g^2 \phi^2\chi^2$. In this model, $\D k\sim k \sim \sqrt{gm\phi_0}$, where $\phi_0$ is the initial inflaton amplitude, and the Hubble radius at the end of the resonance is approximately given by the inflaton mass $m$ \cite{r6}. Setting $\e\sim m$, the order of the relative deviation in the number density in each Hubble volume becomes roughly $(g\phi_0/m)^{-3/4}$. It turns out that a rigorous derivation, which takes into account the details like the cosmological expansion,  agrees with this naive estimate \cite{r5}. For a realistic set of values given in \cite{r6}, namely for  $g=10^{-3}$, $\phi_0=M_p/2$ and $m=10^{-6}M_p$, the order of the relative deviation becomes $10^{-2}$. This indicates that the ``perfect'' homogeneity attained after inflation, which is better than one part in $e^{60}$,  is brokendown on Hubble length scales during reheating.  

\acknowledgments{This work is partially supported by Turkish Academy of Sciences via Young Investigator Award Program (T\"{U}BA-GEB\.{I}P).}

\end{document}